\documentclass[12pt,english,longbibliography,nofootinbib,superscriptaddress,12pt,sort&compress,showkeys]{revtex4-1}
\usepackage{ae,aecompl}
\usepackage[T1]{fontenc}
\usepackage[latin9]{inputenc}
\usepackage[active]{srcltx}
\usepackage{amsmath}
\usepackage{graphicx}

\makeatletter
\usepackage{aecompl}\usepackage{epstopdf}

\usepackage{babel}

\makeatother

\usepackage{babel}
\begin{document}
\title{Stochastic model and kinetic Monte Carlo simulation of solute interactions
with stationary and moving grain boundaries. I. Model formulation
and application to one-dimensional systems}
\author{Y. Mishin}
\address{\noindent Department of Physics and Astronomy, MSN 3F3, George Mason
University, Fairfax, Virginia 22030, USA}
\begin{abstract}
\noindent A simple stochastic model of solute drag by moving grain
boundaries (GBs) is presented. Using a small number of parameters,
the model describes solute interactions with GBs and captures nonlinear
GB dynamics, solute saturation in the segregation atmosphere, and
the breakaway from the atmosphere. The model is solved by kinetic
Monte-Carlo (KMC) simulations with time-dependent transition barriers.
The non-Markovian nature of the KMC process is discussed. In Part
I of this work, the model is applied to planar GBs driven by an external
force. The model reproduces all basic features of the solute drag
effect, including the maximum of the drag force at a critical GB velocity.
The force-velocity functions obtained depart from the scaling predicted
by the classical models by Cahn and Lücke-Stüwe, which are based on
more restrictive assumptions. The paper sets the stage for Part II,
in which the GB will be treated as a 2D solid-on-solid interface. 
\end{abstract}
\maketitle

\section{Introduction}

Many properties of technological materials are controlled by the motion,
or resistance to the motion, of grain boundaries (GBs) under the action
of capillarity and other thermodynamic driving forces \citep{Balluffi95}.
In alloys, the interaction of GBs with alloy components can drastically
reduce the GB mobility due to the solute drag effect. This effect
has been studied experimentally, theoretically, and by computer simulations
for decades. The first quantitative model of the solute drag was proposed
by Cahn \citep{Cahn-1962} and Lücke et al.~\citep{Lucke-Stuwe-1963,Lucke:1971aa}.
Their model predicts a highly nonlinear relation between the GB velocity
and the drag force. The drag force exhibits a maximum separating two
kinetic regimes. In the low-velocity regime, the GB drags the solute
segregation atmosphere. In the high-velocity regime, the boundary
breaks away from the atmosphere but soon forms a new, lighter atmosphere
that poses less resistance to the GB motion. 

On the computational side, the solute drag was studied by the phase
field \citep{Wang03,Li:2009aa,Shahandeh:2012aa,Gronhagen:2007aa,Abdeljawad:2017aa,Alkayyali:2021uo}
and phase-field crystal \citep{Greenwood:2012aa} methods, and by
molecular (MD) simulations \citep{Mendelev02a,Sun:2014aa,Wicaksono:2013aa,Mendelev:2001aa,Rahman:2016aa,Kim:2008aa}.
MD offers the most powerful approach to studying the solute drag.
It provides access to all atomic-level details of the GB motion and
does not rely on any assumptions or approximations other than those
built into the interatomic potential. However, the timescale of MD
simulations is presently limited to about a hundred nanoseconds, which
is too short for reliable modeling of diffusion in the lattice by
the vacancy mechanism. Despite this limitation, recent work \citep{Koju:2020aa,Koju:2021aa}
has shown that the ``short circuit'' GB diffusion and the cloud of
non-equilibrium vacancies surrounding the moving boundary provide
sufficient diffusion mobility to observe the solute drag on the MD
time scale. Significant insights into the solute drag mechanisms were
obtained, especially regarding the role of the in-plane GB diffusion.
Presently, such simulations remain very challenging. They heavily
rely on computational power and depend on the availability and reliability
of interatomic potentials for alloy systems.

In this work, we approach the solute drag problem from a different
direction. Our goal is to create a minimalist model that would be
as simple as possible and only depend on a small number of parameters
but would still capture the most essential physics of the solute drag
effect. The model proposed here is stochastic and is solved by kinetic
Monte Carlo (KMC) simulations. The paper is divided into two parts.
In Part I, we introduce our model and demonstrate its first application.
Section \ref{sec:Kinetic-model} describes the kinetic theory underlying
the model and its KMC implementation. In section \ref{sec:1D-model},
we specialize the model to GB motion in one-dimensional (1D) systems.
We show that the model reproduces all basic features of the solute
drag effect, including the maximum of the force-velocity function
predicted by the classical models \citep{Cahn-1962,Lucke-Stuwe-1963,Lucke:1971aa}.
However, the model departs from the scaling predicted by the classical
models as it captures more realistic GB dynamics. In Part II \citep{Mishin_2023_RW_part_II},
we will present a 2D version of the model, which treats the GB as
a solid-on-solid interface with an adjustable interface energy and
reproduces a GB roughening transition in both stationary and moving
boundaries. This will allow us to study the dynamic roughening phenomenon
and its impact on GB migration mechanisms and the solute drag. 

\section{The kinetic model\label{sec:Kinetic-model}}

\subsection{Random walk without pinning}

\begin{figure}
\begin{centering}
\includegraphics[width=0.5\textwidth]{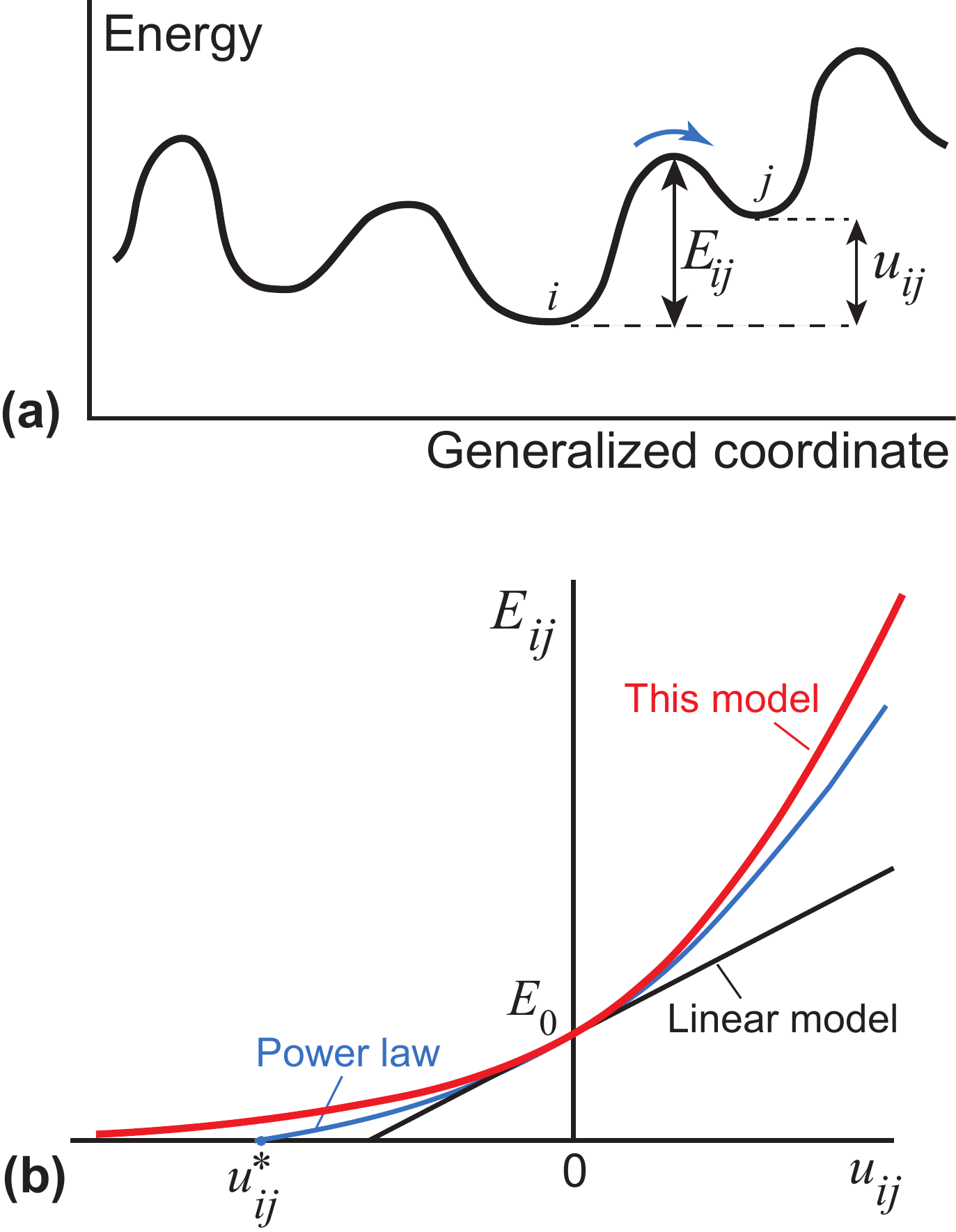}
\par\end{centering}
\caption{(a) Schematic 1D representation of energy landscape of a system capable
of jumping between energy minima by thermal fluctuations. $E_{ij}$
is the energy barrier from state $i$ to state $j$ with energies
$u_{i}$ and $u_{j}$, respectively. (b) Energy barrier as a function
of energy difference $u_{ij}=u_{j}-u_{i}$ in the present model compared
with the linear and power-law models.\label{fig:Schematic-1D}}

\end{figure}

Consider a system whose potential energy surface has a set of local
minima. The system is coupled to a thermostat at a temperature $T$
and can transition between the energy minima by thermal fluctuations
(Fig.~\ref{fig:Schematic-1D}(a)). We further assume that such transitions
(jumps) only occur between states separated by a single energy barrier.
We adopt the harmonic transition state theory (TST) \citep{Vineyard:1957vo},
by which the transition rate $\omega_{ij}$ from state $i$ to state
$j$ is given by 
\begin{equation}
\omega_{ij}=\nu_{0}\exp\left(-\dfrac{E_{ij}}{k_{B}T}\right),\label{eq:1.01}
\end{equation}
where $E_{ij}$ is the transition barrier, $k_{B}$ is Boltzmann's
constant, and $\nu_{0}$ is the attempt frequency. For simplicity,
$\nu_{0}$ is assumed to be the same for all transitions.

The following rule is introduced for the transition barriers:
\begin{equation}
E_{ij}=\begin{cases}
E_{0}\exp\left(\dfrac{u_{ij}}{2E_{0}}\right), & u_{ij}\leq0,\\
u_{ij}+E_{0}\exp\left(-\dfrac{u_{ij}}{2E_{0}}\right), & u_{ij}>0,
\end{cases}\label{eq:1.02}
\end{equation}
where $u_{ij}=u_{j}-u_{i}$ is the energy difference between initial
($u_{i}$) and destination ($u_{j}$) states, and $E_{0}$ is the
barrier between the states of equal energy ($u_{i}=u_{j}$). According
to Eq.(\ref{eq:1.02}), transition barriers to lower (higher) energy
states are lower (higher) than $E_{0}$. Generally, the barrier is
a nonlinear function of $u_{ij}$ (Fig.~\ref{fig:Schematic-1D}(b)).
However, if the energy difference is small, $\left|u_{ij}\right|\ll E_{0}$,
then the barrier can be approximated by $E_{ij}=E_{0}+u_{ij}/2$.
This linear approximation is often employed to describe weakly driven
systems. If the energy difference is large ($\left|u_{ij}\right|\gg E_{0}$),
then the barrier is large and close to $u_{ij}$ for transitions to
higher-energy states and exponentially small for transitions to lower-energy
states. 

Previous energy-barrier models assumed that, under a sufficiently
large energy difference $-u_{ij}$, the barrier is suppressed to zero
at a critical value $u_{ij}^{*}$. It was further assumed that the
barrier follows the power law $E_{ij}\propto\left(u_{ij}-u_{ij}^{*}\right)^{\alpha}$
at $u_{ij}\gtrsim u_{ij}^{*}$ and remains strictly zero at $u_{ij}<u_{ij}^{*}$.
Theoretical models predict the critical exponent $\alpha=3/2$ \citep{Cahn:2001wh,Cottrell:2002ub,Ivanov08a},
although computer simulations often deviate from this value \citep{Chachamovitz:2018vm}.
In the present model, the zero-barrier point is regularized by replacing
the power law with an exponential decay of the barrier with increasing
$\left|u_{ij}\right|$. The exponential decay ensures that the barrier
remains positive under any driving force. This regularization is introduced
for computational convenience and does not affect any physically meaningful
results. Indeed, the TST underlying Eq.(\ref{eq:1.01}) is only valid
when $E_{ij}\gg k_{B}T>0$. Any results obtained at $E_{ij}\rightarrow0$
lie outside the validity domain of the TST and must be ignored in
all applications of this model.

If the system is confined to a finite-size domain in the configuration
space, it eventually reaches a state of dynamic equilibrium. Note
that the energy-barrier relation (\ref{eq:1.02}) satisfies the detailed
balance condition
\begin{equation}
\omega_{ij}\exp\left(-\dfrac{u_{i}}{k_{B}T}\right)=\omega_{ji}\exp\left(-\dfrac{u_{j}}{k_{B}T}\right),\label{eq:1.03}
\end{equation}
where the exponential terms represent the Boltzmann probabilities
of finding the system in the respective states. On the other hand,
an open system subjected to a driving force executes a driven random
walk and never reaches equilibrium. This is illustrated in Figure
\ref{fig:-Schematic-barrier} by a 1D example in which a uniform driving
force $F>0$ tilts the periodic energy landscape of the system and
changes the initial barrier $E_{0}$ to 
\begin{equation}
E^{(+)}=E_{0}\exp\left(-\dfrac{Fa}{2E_{0}}\right)\label{eq:1.03a}
\end{equation}
for forward jumps and 
\begin{equation}
E^{(-)}=Fa+E_{0}\exp\left(-\dfrac{Fa}{2E_{0}}\right)\label{eq:1.03b}
\end{equation}
for backward jumps ($a$ being the energy period). The bias between
the forward and backward barriers causes a drift of the system in
the force direction with the velocity
\begin{equation}
v=a\nu_{0}\left[\exp\left(-\dfrac{E^{(+)}}{k_{B}T}\right)-\exp\left(-\dfrac{E^{(-)}}{k_{B}T}\right)\right].\label{eq:1.04}
\end{equation}

The system evolution can be modeled by KMC simulations implementing
a sequence of jump attempts that may or may not be successful. We
assume for simplicity that each state has the same number $m$ of
available escape routes. At each step of the KMC process, a random
number $r_{1}\in[0,1)$ chooses one of the possible jumps $i\rightarrow j$
from the current state $i$ with equal probability. Then another random
number $r_{2}\in[0,1)$ decides if the chosen jump attempt is successful.
The jump is implemented if $r_{2}<p_{ij}$, where 
\begin{equation}
p_{ij}=\exp\left(-\dfrac{E_{ij}}{k_{B}T}\right)\label{eq:1.05}
\end{equation}
is the success probability; otherwise the attempt fails. In either
case, the clock is advanced by $\Delta t=\left(m\nu_{0}\right)^{-1}$
and the process repeats. $m$ KMC steps correspond to one physical
attempt with the frequency $\nu_{0}$.\footnote{The reader should not confuse the physical transition attempts occurring
with the frequency $\nu_{0}$ and the KMC attempts with the frequency
$m\nu_{0}$. The distinction must be clear from the context. }

\subsection{The pinning effect}

The previous discussion assumed that the energy landscape of the system
and the unbiased barrier $E_{0}$ were fixed. We will now modify this
assumption. Let us first consider unbiased jumps ($u_{i}=u_{j}$).
After each unsuccessful attempt, we will penalize the system by increasing
the jump barriers for all escape routes. After $n$ unsuccessful attempts,
the barriers become
\begin{equation}
E_{t}=E_{0}\left(1+(\alpha-1)\dfrac{\sqrt{t/t_{p}}}{1+\sqrt{t/t_{p}}}\right),\label{eq:activation_with_pinning-1}
\end{equation}
where $t=n\Delta t$ is the discrete time variable, and $\alpha>1$
and $t_{p}>0$ are model parameters. The first attempt ($t=0$) uses
the unpenalized barriers $E_{0}$. If this attempt fails, the barriers
for the second attempt become $E_{1}>E_{0}$. If the second attempt
is also unsuccessful, the barriers become $E_{2}>E_{1}$, and so on.
As long as $t\ll t_{p}$, the barriers grow with time as $\sqrt{t}$.
In the limit of $t\gg t_{p}$, the barriers plateau at 
\begin{equation}
E_{\infty}=\alpha E_{0}>E_{0}.\label{eq:max_barrier-1}
\end{equation}
After the system finally makes a successful jump, the attempt counter
$n$ is reset to zero and the process repeats from the new state. 

In the presence of energy gradients ($u_{i}\neq u_{j}$), the jump
barriers are given by Eq.(\ref{eq:1.02}) with $E_{0}$ replaced by
$E_{t}$ from Eq.(\ref{eq:activation_with_pinning-1}). 

The physical motivation for introducing the time-dependent barriers
is to describe the GB interaction with solute atoms, including the
solute drag effect \citep{Cahn-1962,Lucke-Stuwe-1963,Lucke:1971aa}.
The amount of solute transported to the boundary by diffusion initially
increases as the square root of time, reflecting diffusion kinetics.
The solute atoms form a segregation atmosphere that reduces the boundary
mobility by raising the energy barriers for its random displacements.
After the segregation atmosphere has reached its maximum capacity
(saturation), the barriers remain constant. This behavior is captured
by Eq.(\ref{eq:activation_with_pinning-1}), which ensures that $E_{t}$
grows as $\sqrt{t}$ at $t\ll t_{p}$ and reaches a plateau value
$E_{\infty}$ in the long-time limit ($t\gg t_{p}$).

In addition to the solute drag by moving GBs, this model is relevant
to the motion of other crystalline defects in the presence of segregating
chemical components reducing the defect mobility. The defect in question
can be a lattice dislocation whose mobility is slowed down by a Cottrell
atmosphere of solute atoms \citep{Hirth,Cottrell:1948aa,Cottrell_Bilby,Cottrell:1953aa}.
As another example, consider diffusion of slow impurity atoms in the
presence of highly mobile atoms of another chemical component that
interacts with the impurity atoms creating a short-range order around
them. This short-range order can be treated as a segregation atmosphere
reducing the impurity mobility.

For brevity, the increase in the jump barriers with time will be referred
to as the ``pinning'' effect. This term should not be understood as
literally pinning the system in place. It only refers to the retardation
of the system dynamics caused by the diffusion-controlled formation
of a segregation atmosphere. Accordingly, the parameter $t_{p}$ will
be called the pinning time. The latter is related to the solute diffusion
coefficient $D$ by
\begin{equation}
D=\dfrac{a^{2}}{t_{p}}=\dfrac{t_{0}}{t_{p}}D_{0}.\label{eq:1.06}
\end{equation}
Here, $a$ is a characteristic jump length, $D_{0}=a^{2}/t_{0}$ is
the intrinsic diffusivity of the unpinned system executing a random
walk by thermal fluctuations, and 
\begin{equation}
t_{0}=\dfrac{1}{m\nu_{0}}\exp\left(\dfrac{E_{0}}{k_{B}T}\right)\label{eq:1.06a}
\end{equation}
is the unpinned and unbiased residence time of the system. Likewise,
$E_{\infty}$ has the meaning of the fully pinned jump barrier. The
difference $(E_{0}-E_{\infty})=(1-\alpha)E_{0}<0$ can be interpreted
as the solute binding energy to the system. Accordingly, $\alpha$
will be called the pinning factor of the solute atoms.

Note the simplicity of the proposed model. The input information consists
of five parameters: the jump length $a$, the TST parameters $\nu_{0}$
and $E_{0}$, and the pinning parameters $t_{p}$ and $\alpha$. All
other variables mentioned above, such as $t_{0}$, $E_{\infty}$,
and $D$, can be expressed through the five independent parameters
$(a,\nu_{0},E_{0},t_{p},\alpha)$. With this limited input, the model
can be solved by KMC simulations to describe the dynamics of the driven
system. This simple model captures the main physics of the solute
drag effect as will be demonstrated later in the paper and in Part
II \citep{Mishin_2023_RW_part_II}.

\section{1D model of solute drag\label{sec:1D-model}}

\subsection{GB random walk without pinning}

\begin{figure}
\begin{centering}
\includegraphics[width=0.8\columnwidth]{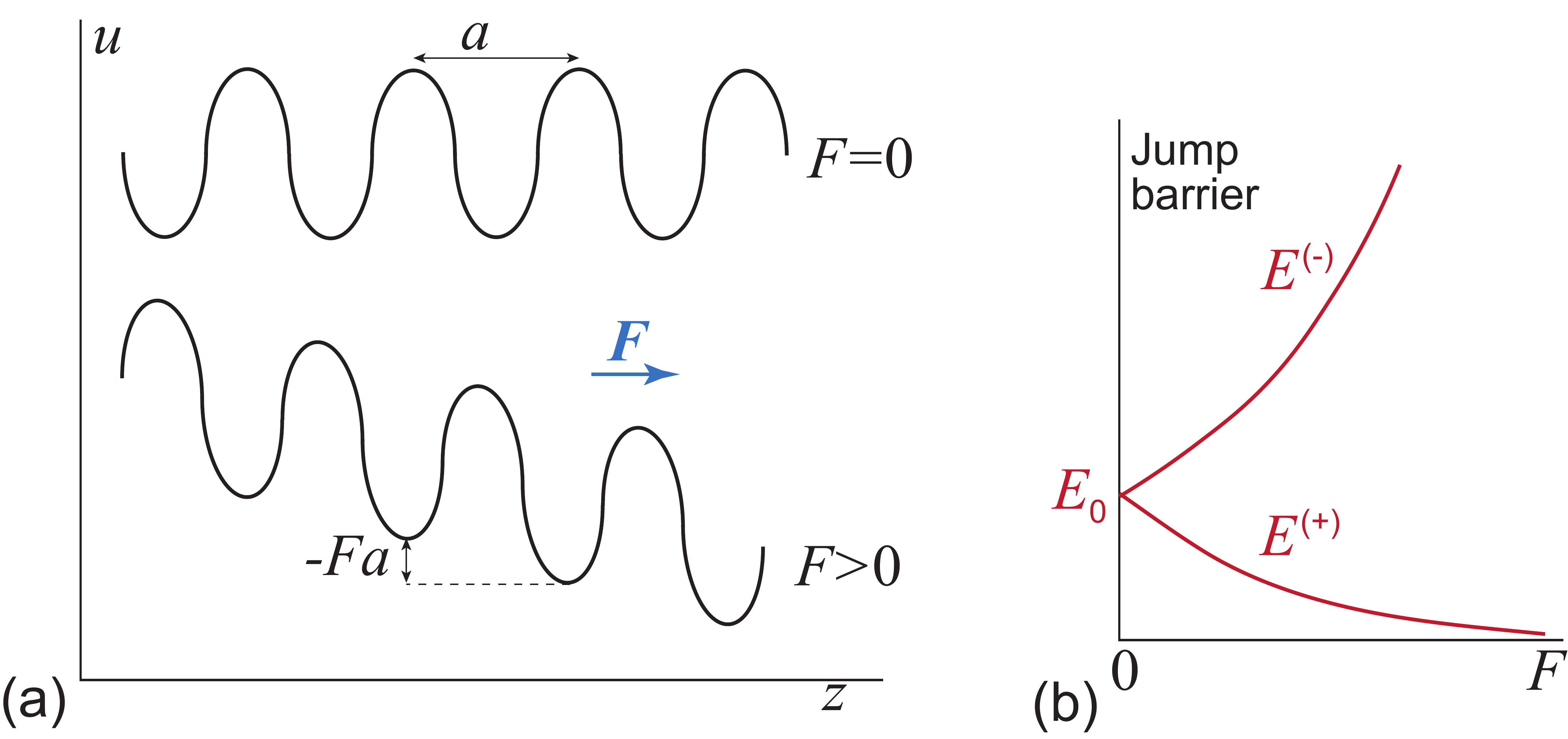}
\par\end{centering}
\caption{(a) Schematic energy landscape of a 1D periodic system before and
after application of a spatially uniform external force $F>0$. (b)
The force suppresses the energy barrier $E^{(+)}$ for forward jumps
and raises the barrier $E^{(-)}$ for backward jumps relative to the
unbiased barrier $E_{0}$.\label{fig:-Schematic-barrier}}
\end{figure}

As the first application, we will consider 1D random walk of a system
on the $z$ axis as depicted in Fig.~\ref{fig:-Schematic-barrier}.
The energy landscape is periodic with a period $a$. This model represents
a planar GB driven by an external force. Note that the classical models
by Cahn and Lücke \citep{Cahn-1962,Lucke-Stuwe-1963,Lucke:1971aa}
also treat the GB as a planar interface and are essentially 1D models.

In the absence of driving forces and pinning, the GB executes an unbiased
random walk with the jump length $a$ and the barrier $E_{0}$. The
escape probability per physical attempt is
\begin{equation}
p_{0}=2\exp\left(-\dfrac{E_{0}}{k_{B}T}\right),\label{eq:1}
\end{equation}
where the factor of two takes into account that the GB can escape
by either a forward or a backward jump ($m=2$). The residence time
of the GB is a discrete stochastic variable $n\nu_{0}^{-1}$, where
$n=1,2,...$ is a counter of attempts. Since the attempts are statistically
independent, the escape probability after $n$ unsuccessful attempts
follows the geometric distribution 
\begin{equation}
\mathsf{P}(n)=p_{0}(1-p_{0})^{n}.\label{eq:geom_distr}
\end{equation}
In the long-time limit ($n\gg1/p_{0}$), this distribution becomes
exponential, 
\begin{equation}
\mathsf{P}(n)=p_{0}e^{-p_{0}n}.\label{eq:exp_distrib}
\end{equation}
It can be shown that the expectation value of the residence time is
\begin{equation}
t_{0}=\dfrac{1}{\nu_{0}p_{0}}=\dfrac{1}{2\nu_{0}}\exp\left(\dfrac{E_{0}}{k_{B}T}\right).\label{eq:t_av}
\end{equation}

Now suppose a driving force $F>0$ is applied to the GB. The force
reduces the forward jump barrier and raises the backward jump barrier
according to Eqs.(\ref{eq:1.03a})-(\ref{eq:1.03b}); see also Figure
\ref{fig:-Schematic-barrier}. This bias causes a drift of the GB
with the velocity given by Eq.(\ref{eq:1.04}). When the force is
small ($F\ll E_{0}/a$), the barriers are approximately linear in
the force, $E^{(\pm)}=E_{0}\mp Fa/2$, and Eq.(\ref{eq:1.04}) predicts
the linear dynamics
\begin{equation}
v=MF,\label{eq:1.07}
\end{equation}
where
\begin{equation}
M=\dfrac{a^{2}\nu_{0}}{k_{B}T}\exp\left(-\dfrac{E_{0}}{k_{B}T}\right)\label{eq:1.08}
\end{equation}
is the GB mobility. A medium force ($F\approx E_{0}/a$) causes an
upward deviation from the linear law. In the large-force limit ($F\gg E_{0}/a$),
the velocity slows down and follows the asymptotic relation
\begin{equation}
v=a\nu_{0}\exp\left(-{\displaystyle \dfrac{E_{0}\exp\left(-\dfrac{Fa}{2E_{0}}\right)}{k_{B}T}}\right).\label{eq:1.04-1}
\end{equation}
The upper bound of the GB velocity is $a\nu_{0}$ but this velocity
is never reached because the barrier never becomes strictly zero.
The physical motivation for preventing a zero barrier is that the
GB jumps are accompanied by energy dissipation in the form of phonon
drag and, in alloys, the solute drag. The model attempts to capture
the dissipation effects by keeping the barrier positive even under
a strong force. It should also be noted that Eq.(\ref{eq:1.04-1})
is physically meaningful only as long as the numerator in the exponent
is $\gg k_{B}T$; otherwise the TST cannot be applied. 

\subsection{GB random walk with pinning}

We next consider unbiased ($F=0$) GB walk with pinning. Unsuccessful
jump attempts are now penalized according to Eq.(\ref{eq:activation_with_pinning-1}).
For a fully pinned GB ($E_{t}\approx E_{\infty}$), the escape probability
per physical attempt is 
\begin{equation}
p_{\infty}=2\exp\left(-\dfrac{E_{\infty}}{k_{B}T}\right),\label{eq:p_infinity}
\end{equation}
and the number of failed attempts follows the geometric distribution
\begin{equation}
\mathsf{P}(n)=p_{\infty}(1-p_{\infty})^{n}.\label{eq:geom_distr-1}
\end{equation}
In the large-$n$ limit, this distribution converges to exponential,
\begin{equation}
\mathsf{P}(n)=p_{\infty}e^{-p_{\infty}n},\label{eq:exp_distrib-1}
\end{equation}
and the expectation value of the residence time becomes
\begin{equation}
t_{\infty}=\dfrac{1}{\nu_{0}p_{\infty}}=t_{0}\exp\left(\dfrac{E_{\infty}-E_{0}}{k_{B}T}\right)=t_{0}\exp\left(\dfrac{(\alpha-1)E_{0}}{k_{B}T}\right).\label{eq:t_av-1}
\end{equation}

If the pinning time is long ($t_{p}\gg t_{0}$), the atmosphere formation
is a slow process. Then there is a high probability that the GB escapes
before any significant atmosphere can form. The pinning has little
effect on the GB walk. This is the case for slow solute diffusion
($D\ll D_{0}$). If $t_{p}\ll t_{0}$, a fully saturated atmosphere
forms before the GB has a chance to escape. Accordingly, the jump
barrier is close to $E_{\infty}$ and the residence time is close
to $t_{\infty}$. This is the case when the solute diffusion is fast
($D\gg D_{0}$). On the intermediate time scale between the two limits
($t_{p}\approx t_{0}$), the residence time no longer follows the
geometric or exponential distribution, making the process non-Markovian.
The expectation value of the residence time lies between $t_{0}$
and $t_{\infty}$. We call this kinetic regime ``active pinning''.

We next apply a driving force $F>0$ causing the GB to drift in the
positive $z$ direction. This drift cannot be described analytically
and was studied by KMC simulations. The simulations were performed
in dimensionless variables using $a$, $\nu_{0}^{-1}$ and $k_{B}T$
as the units of length, time, and energy, respectively. All KMC results
reported below are for $t_{0}\nu_{0}=50$ and thus $E_{0}/k_{B}T=4.6$.
These values were chosen as a compromise between computational efficiency
and the $E_{0}/k_{B}T\gg1$ requirement of the TST.

\begin{figure}
\textbf{(a)} \includegraphics[width=0.7\textwidth]{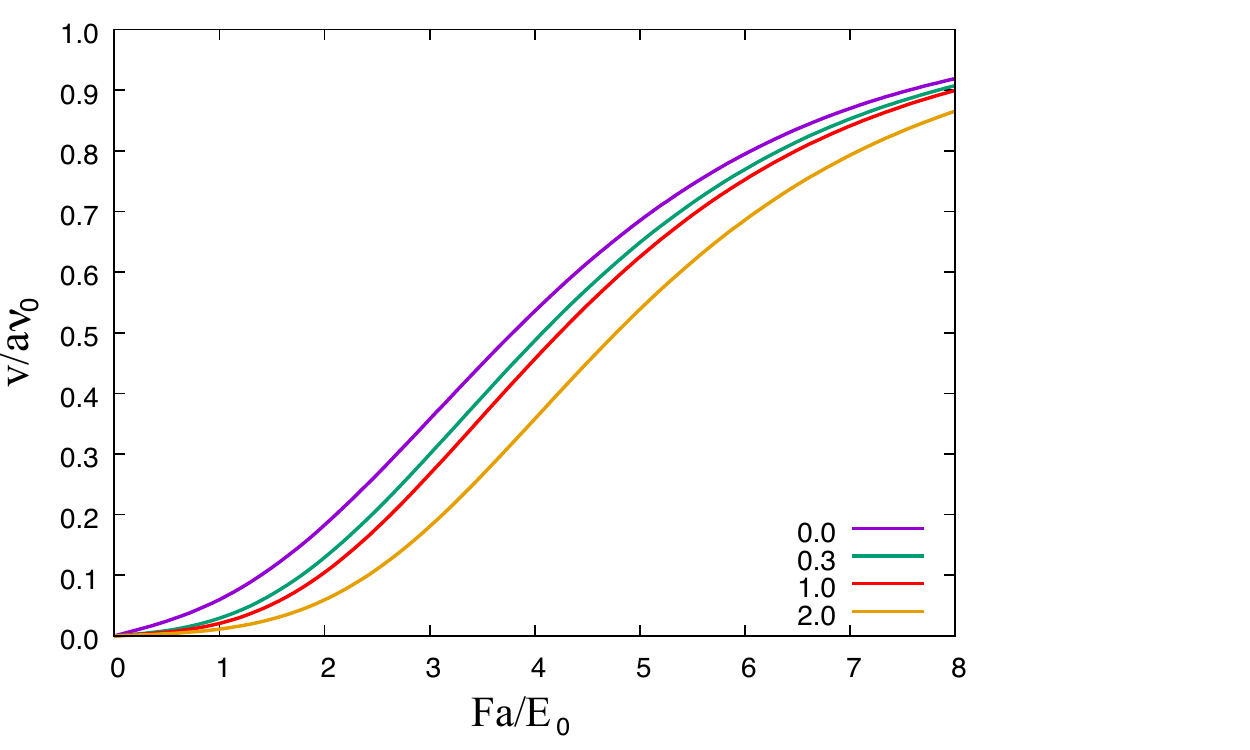}

\bigskip{}
\bigskip{}

\textbf{(b)} \includegraphics[width=0.7\textwidth]{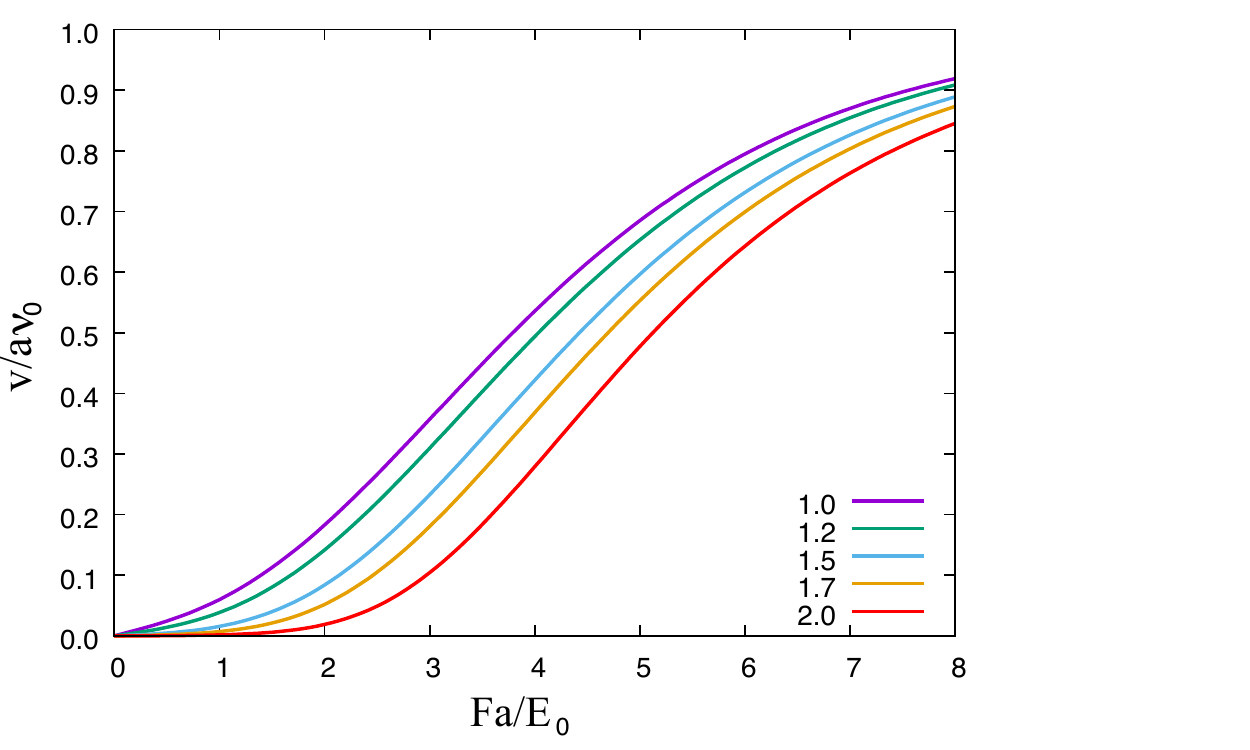}

\caption{Velocity-force relations for: (a) several normalized solute diffusivities
$D/D_{0}$ indicated in the key at a fixed pinning factor $\alpha=1.5$;
(b) several $\alpha$ values indicated in the key at a fixed $D/D_{0}=2.0$.\label{fig:Velocity-force-1D}}
\end{figure}

Figure \ref{fig:Velocity-force-1D}(a) shows the velocity-force functions
for a series of normalized solute diffusivities $D/D_{0}$ with a
fixed pinning factor $\alpha=1.5$. Figure \ref{fig:Velocity-force-1D}(b)
shows such functions for a series of $\alpha$ values with a fixed
$D/D_{0}=2.0$. As expected, the results for $D/D_{0}=0$ (no solute
diffusion) and $\alpha=1$ (no solute segregation) perfectly match
the analytical solution (\ref{eq:1.04}) (not shown in the figure).
The plots demonstrate that increasing the solute diffusivity and/or
the pinning factor reduces the GB velocity under a given driving force,
which is a manifestation of the solute drag effect.

\begin{figure}
\textbf{(a)} \includegraphics[width=0.42\textwidth]{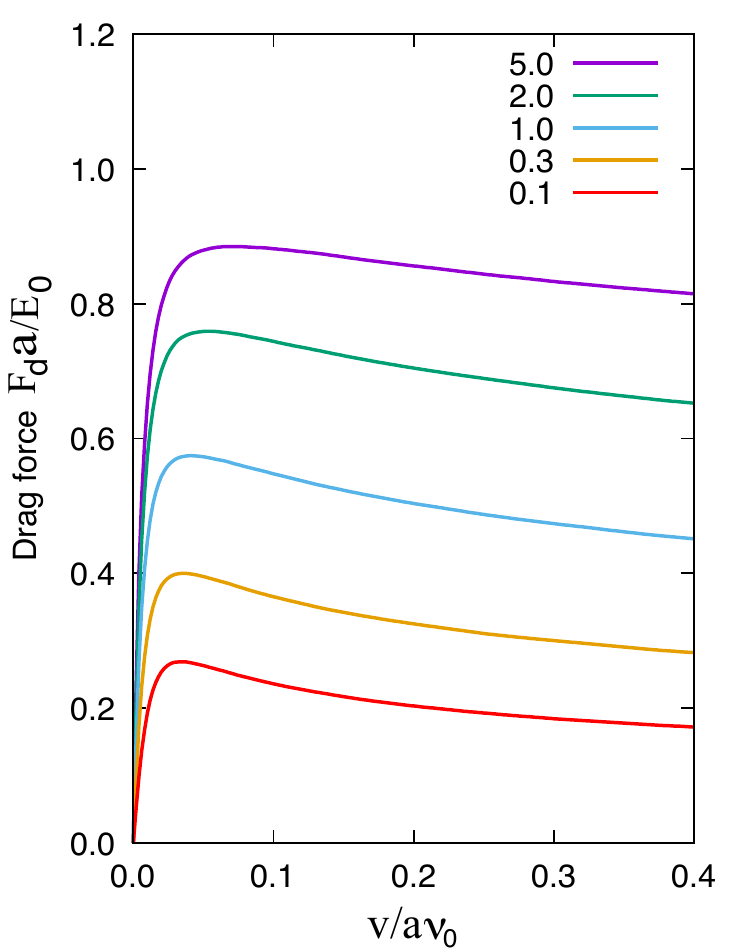}
\textbf{\enskip{}(b)} \includegraphics[width=0.42\columnwidth]{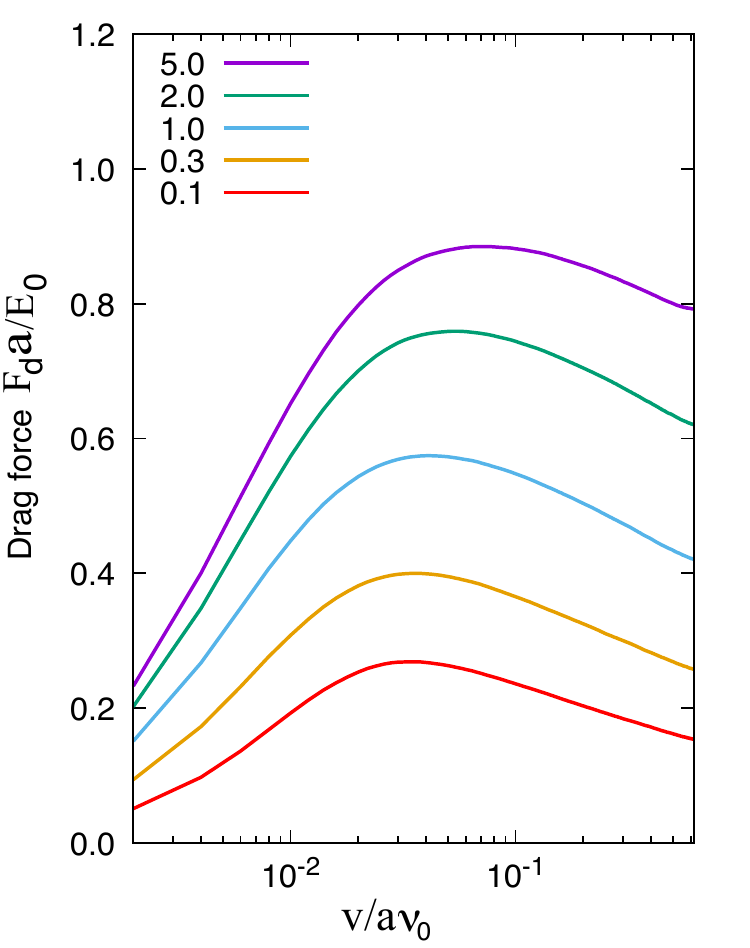}

\bigskip{}
\bigskip{}

\textbf{(c)} \includegraphics[width=0.42\textwidth]{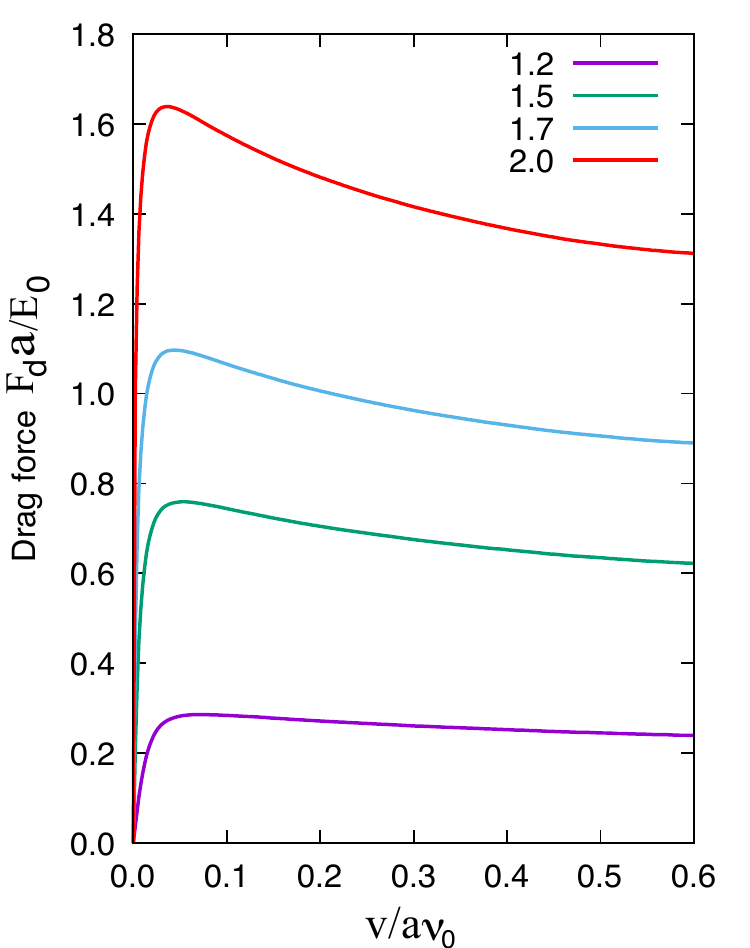}
\textbf{\enskip{}(d)} \includegraphics[width=0.42\columnwidth]{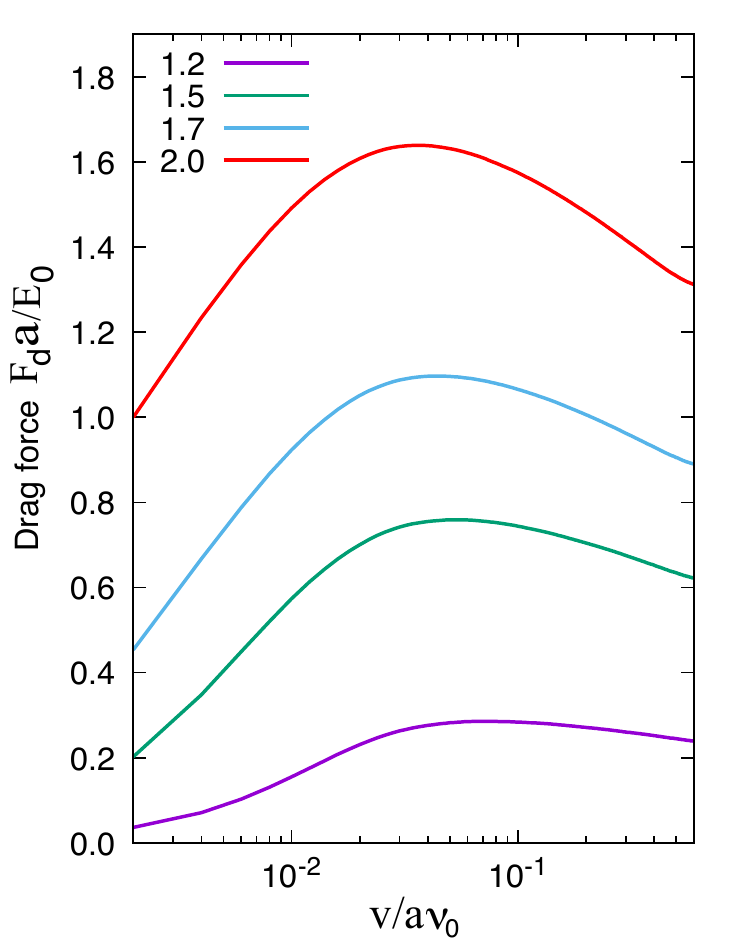}

\caption{Solute drag force as a function of velocity for: (a,b) several normalized
solute diffusivities $D/D_{0}$ indicted in the key at a fixed pinning
factor $\alpha=1.5$; (c,d) several $\alpha$ values indicted in the
key at a fixed $D/D_{0}=2.0$. Panels (b) and (d) use the logarithmic
velocity scale to better reveal the drag-breakaway transition.\label{fig:Solute-drag-1D}}
\end{figure}

The solute drag force $F_{d}$ is the difference between the force
required to drive a segregated GB and the force to drive an unpinned
GB ($D/D_{0}=0$ or $\alpha=1$) with the same velocity. The velocity
dependence of the normalized solute drag force, $F_{d}a/E_{0}$, is
displayed in Figs.~\ref{fig:Solute-drag-1D}(a,b) for several $D/D_{0}$
values at a fixed $\alpha=1.5$, and in Figs.~\ref{fig:Solute-drag-1D}(c,d)
for several $\alpha$ values at a fixed $D/D_{0}=2.0$. In agreement
with the classical models, the drag force reaches a maximum at a critical
velocity $v_{*}$ separating the solute drag regime at $v<v_{*}$
and the breakaway regime at $v>v_{*}$. The transition between the
two regimes occurs continuously over a wide velocity range. This transition
is best revealed using the logarithmic velocity axis as in Figs.~\ref{fig:Solute-drag-1D}(b,d).

Although the observation of the two kinetic regimes is in qualitative
agreement with the classical models \citep{Cahn-1962,Lucke-Stuwe-1963,Lucke:1971aa},
there are also significant differences. For example, Cahn's model
\citep{Cahn-1962} predicts that the drag force is a function of the
dimensionless parameter $av/D$ {[}Eqs.(16)-(8) in \citep{Cahn-1962}{]}.
According to this prediction, the maximum drag force must be independent
of the solute diffusivity $D$, while the peak position $v_{*}$ must
be proportional to $D$. In our model, the solute diffusivity $D$
is given by Eq.(\ref{eq:1.06}), so Cahn's scaling variable is $vt_{p}/a$.
This variable has the meaning of the distance traveled by the moving
GB during the pinning time $t_{p}$. Contrary to this prediction,
the peak force obtained by the simulations sharply increases with
$D$ (Figs.~\ref{fig:Solute-drag-1D}(a,b)). The peak velocity $v_{*}$
is not proportional to $D$ either. A similar lack of the $av/D$
scaling was observed in previous KMC simulations within a 2D Ising
model \citep{Mendelev:2001aa} and 3D solid-on-solid model \citep{Wicaksono:2013aa}.
This discrepancy is due to the crude approximations underlying the
classical models. Both our present model and Cahn's theory \citep{Cahn-1962}
represent the GB by a planar interface, but our model captures the
solute saturation effect missing in Cahn's theory and explicitly treats
the nonlinear GB dynamics both with and without the GB-solute interactions. 

Figure \ref{fig:Solute-drag-1D} also shows the trend for the drag
force maximum to widen with increasing solute diffusivity and/or decreasing
pinning factor. Although not shown in Fig.~\ref{fig:Solute-drag-1D},
at sufficiently large $D/D_{0}$ values and/or sufficiently small
$\alpha$, the maximum smooths out. The thinning of the segregation
atmosphere with increasing velocity becomes a continuous process not
accompanied by a breakaway event. 

\section{Discussion and conclusions}

This work aimed to develop a minimalist model capturing the main physics
of the solute drag by moving GBs. The key feature of the solute drag
effect is the kinetic competition between GB migration and diffusion
of the solute atoms. A moving GB, driven by an external force, tries
to increase its mobility by breaking away from the solute segregation
atmosphere. The formation of the atmosphere is kinetically controlled
by solute diffusion. If the latter outpaces the GB migration, a heavy
atmosphere forms that slows the GB down. If the GB mobility is high
relative to solute diffusion, the GB only carries a light atmosphere
and can move faster. 

The model proposed here describes this kinetic competition. It represents
both linear and nonlinear GB dynamics using only three parameters:
$a$, $\nu_{0}$ and $E_{0}$. The solute interaction with the GB
and the solute diffusivity are represented by two more parameters:
the pinning strength $\alpha$ and the pinning time $t_{p}$. The
solute diffusion is included in the model through the square root
time dependence of the GB jump barriers. Out of the five parameters
mentioned, $\nu_{0}$ and $a$ set the time and length scales of the
problem and are unrelated to the competition of the kinetic regimes.
The key parameters of the model are $E_{0}$, $\alpha$ and $t_{p}$.\footnote{The 2D version of the model presented in \citep{Mishin_2023_RW_part_II}
additionally includes the GB energy as another parameter. This parameter
controls the GB migration mechanisms and capillary fluctuations at
high temperatures.} We believe that this model achieves the ultimate simplicity in describing
the solute drag effect. Nothing in the model can be removed without
losing the underlying physics. 

The 1D version of the model presented in this paper reproduces the
main features of the solute drag, including the drag force maximum
at a critical velocity. The model predictions are in qualitative agreement
with the classical models \citep{Cahn-1962,Lucke-Stuwe-1963,Lucke:1971aa},
which are also based on 1D geometry. However, the classical models
rely on more restrictive assumptions, such as the dilute solution
approximation and linear GB dynamics in the absence of solute atoms.
The present model captures some of the missing features, including
nonlinear dynamics and the solute saturation effect. 

The introduction of time-dependent transition barriers in this model
raises some theoretical questions that are not apparent in the 1D
version of the model but are more relevant to the 2D version \citep{Mishin_2023_RW_part_II}
and other possible applications. One of the questions is whether the
KMC simulations based on this model implement a Markov chain. On one
hand, the GB jump probabilities from a given state are statistically
independent of the previous jumps, as in a Markov chain. On the other
hand, the residence time probability distribution is not exponential
as it should be in a continuous-time Markov process, making our process
non-Markovian. Specifically, the random walk with pinning introduced
in this model can be classified as a homogeneous semi-Markov process
\citep{Chari:1994aa,Yu:2010aa}. The homogeneity means that the residence
time distribution depends only on the time counted from the arrival
at the current state, not the absolute time. The TST requirement of
relatively high escape barriers implies long residence times with
many unsuccessful attempts. As such, the random walk can be treated
as a continuous-time process \citep{Yu:2010aa}. However, in the actual
simulations, the residence time cannot be too long for computational
reasons. In some cases where the average number of failed attempts
is not too large, the simulations implement a discrete-time semi-Markov
process \citep{Yu:2010aa}. 

Mathematical analysis of random walk with pinning is beyond the scope
of this work. We are more concerned with the consequences of the non-Markovian
nature of the process for the physical behavior of the system. One
question is whether the random walk always converges to a unique steady
state. While we cannot present a general proof that it always does,
in all cases tested in this work, the KMC simulations did converge
to a steady state that was independent of the initial condition. This
was found for both driven processes as well as stationary states arising
in a bound system in the absence of external forces. The steady-state
occupation probabilities do not generally follow the Boltzmann distribution.
The detailed balance condition in the form of Eq.(\ref{eq:1.03})
is not satisfied. However, it is accurately followed when the non-Boltzmann
occupation probabilities are used to formulate the detailed balance.
Given that the residence time does not correlate with the jump directions,
the microscopic reversibility is also preserved \citep{Wang:2007aa}.
Some of these features are illustrated by a simple three-level model
with pinning presented in the Appendix. 

Transitions between different states of GBs, dislocations, and other
crystalline defects subject to active pinning are intrinsically non-Markovian,
whether the system is driven by an applied force or fluctuates around
a fixed average position. At best, the chains of such transitions
constitute semi-Markov processes \citep{Maes:2009aa}. More details
related to simulations of pinned systems will be discussed in Part
II of this work \citep{Mishin_2023_RW_part_II}.

\bigskip{}

\noindent \textbf{Acknowledgements} 

This research was supported by the National Science Foundation, Division
of Materials Research, under Award no. 2103431.

\section*{Appendix: three-level system with pinning}

\noindent In this Appendix, we present a toy model that illustrates
some of the features of random walk in the presence of pinning.

Consider a three-level system coupled to a thermostat and subject
to the pinning effect introduced in the main text. The system can
spontaneously jump between the states starting from some initial condition.
We can model this process by a KMC simulation. At each KMC step, two
random numbers, $r_{1}$ and $r_{2}$, are generated in a unit interval.
$r_{1}$ selects one of the two states, $j$, different from the current
state $i$ with equal probability. Then $r_{2}$ decides if the jump
$i\rightarrow j$ is implemented, depending on whether $r_{2}<p_{ij}$
(successful attempt) or $r_{2}\geq p_{ij}$ (failed attempt). Here,
\begin{equation}
p_{ij}=e^{-E_{ij}/\theta}\label{eq:s1}
\end{equation}
is the jump probability, $\theta$ is reduced temperature, and $E_{ij}$
is the $i\rightarrow j$ jump barrier. $E_{ij}$ depends on the state
energies $u_{i}$ and $u_{j}$:
\begin{equation}
E_{ij}=\begin{cases}
E_{t}\exp\left(\dfrac{u_{ij}}{2E_{t}}\right), & u_{ij}\leq0,\\
u_{ij}+E_{t}\exp\left(-\dfrac{u_{ij}}{2E_{t}}\right), & u_{ij}>0,
\end{cases}\label{eq:s2}
\end{equation}
where $u_{ij}=u_{j}-u_{i}$ and $E_{t}$ is the unbiased (when $u_{i}=u_{j}$)
jump barrier. The latter is given by
\begin{equation}
E_{t}=1+(\alpha-1)\dfrac{\sqrt{\tau/\tau_{p}}}{1+\sqrt{\tau/\tau_{p}}},\label{eq:s3}
\end{equation}
where $\alpha>1$ is the pinning factor, $\tau_{p}$ is the pinning
time, and $\tau$ is the elapsed time after the previous jump. In
the KMC simulations, $\tau$ is a discrete variable equal to the number
of previously failed attempts. After each successful jump, $\tau$
is reset to zero. 

Note that $p_{ij}$ and $p_{ji}$ satisfy the equation
\begin{equation}
e^{-u_{i}/\theta}p_{ij}=e^{-u_{j}/\theta}p_{ji}.\label{eq:s3b}
\end{equation}
This equation looks like a detailed balance relation with Boltzmann's
occupation probabilities. However, it cannot be interpreted this way
because $p_{ij}$ and $p_{ji}$ are independently fluctuating variables
corresponding to generally different \emph{$E_{t}$} values. Averaging
over a long KMC trajectory is required for testing the detailed balance
hypothesis, which will be done below.

Let us first consider two limiting cases. Suppose $\tau_{p}\gg\tau_{0}$,
where $\tau_{0}$ is the unpinned and unbiased residence time, 
\begin{equation}
\tau_{0}=\dfrac{1}{2}e^{1/\theta}.\label{eq:s4}
\end{equation}
Then Eq.(\ref{eq:s3}) gives $E_{t}=1$ and the pinning effect is
negligible. In the other limit, when $\tau_{p}\ll\tau_{0}$, $E_{t}$
increases with time in proportion to $\sqrt{\tau}$ to mimic the diffusion-controlled
kinetics of the pinning process. In the $\tau_{p}/\tau_{0}\rightarrow0$
limit, $E_{t}$ tends to $E_{\infty}=\alpha$; the system gets pinned
instantly and continues to evolve with the barrier $E_{t}=E_{\infty}>1$. 

In both limiting cases, the unbiased barrier $E_{t}$ is time-independent
and the random walk between the states is a Markov chain. Accordingly,
Eq.(\ref{eq:s3b}) is a true detailed balance relation with Boltzmann's
occupation probabilities. Furthermore, the unpinned and fully pinned
systems must converge to the same steady state with Boltzmann's occupation
probabilities 
\begin{equation}
c_{i}^{B}=\dfrac{1}{\mathcal{P}}e^{-u_{i}/\theta},\label{eq:s6}
\end{equation}
where 
\begin{equation}
\mathcal{P}=\sum_{i}e^{-u_{i}/\theta}\label{eq:s7}
\end{equation}
is the partition function. The ensemble-averaged system energy is
then
\begin{equation}
\varepsilon_{B}=\sum_{i}e^{-u_{i}/\theta}u_{i}\label{eq:s8}
\end{equation}
and the heat capacity is
\begin{equation}
C_{B}=\dfrac{d\varepsilon_{B}}{d\theta}=\dfrac{1}{\theta^{2}}\left(\dfrac{1}{\mathcal{P}}\sum_{i}u_{i}^{2}e^{-u_{i}/\theta}-\varepsilon^{2}\right).\label{eq:s9}
\end{equation}

Between the two extremes lies the case of active pinning in which
$\tau_{p}\approx\tau_{0}$. The unbiased barrier $E_{t}$ is then
stochastic and time-dependent, making the random walk a semi-Markov
process. Analytical treatment of this case is challenging but we can
study it by KMC simulations. The questions we seek to answer are:
\begin{itemize}
\item Do the simulations converge to a steady state, and if so, does the
steady state depend on the initial condition?
\item What are the steady-state occupation probabilities $c_{i}$ of the
energy levels? Generally, they need not follow Boltzmann's distribution
(\ref{eq:s6}).
\item Do the energy fluctuations in the steady state follow the canonical
distribution \citep{Landau-Lifshitz-Stat-phys,Mishin:2015ab}?
\item When the system is in a steady state, do the jumps satisfy the detailed
balance condition or only the general balance condition \citep{Manousiouthakis:1999wn}?
\end{itemize}
The last question requires a clarification. If the system reaches
a steady state, it must satisfy at least the general balance condition
\citep{Manousiouthakis:1999wn}
\begin{equation}
\sum_{j\neq i}J_{ij}=\sum_{j\neq i}J_{ji}\qquad(\mathrm{fixed\ }i)\label{eq:s10}
\end{equation}
for each state $i.$ Here, $J_{ij}$ is the $i\rightarrow j$ jump
rate (number of $i\rightarrow j$ jumps per unit time) averaged over
a long KMC trajectory. Equation (\ref{eq:s10}) states that the jumps
in and out of any state $i$ balance each other so that the occupation
probability $c_{i}$ is time-independent. The question is whether
the detailed balance relations
\begin{equation}
J_{ij}=J_{ji}\label{eq:s5}
\end{equation}
hold for all \emph{individual} $(i,j)$ pairs, which is obviously
a stronger condition than Eq.(\ref{eq:s10}).

It should be reminded that this model only makes physical sense when
$\theta\ll1$; otherwise the transition state theory underlying Eq.(\ref{eq:s1})
is invalid. We performed KMC simulations at temperatures $0.1<\theta<0.35$.
Here, the upper bound attempts to meet the TST requirement while the
lower bound is imposed by the computational challenge of working with
high barriers. The simulation results are summarized below.

\begin{figure}
\textbf{(a)} \includegraphics[width=0.43\textwidth]{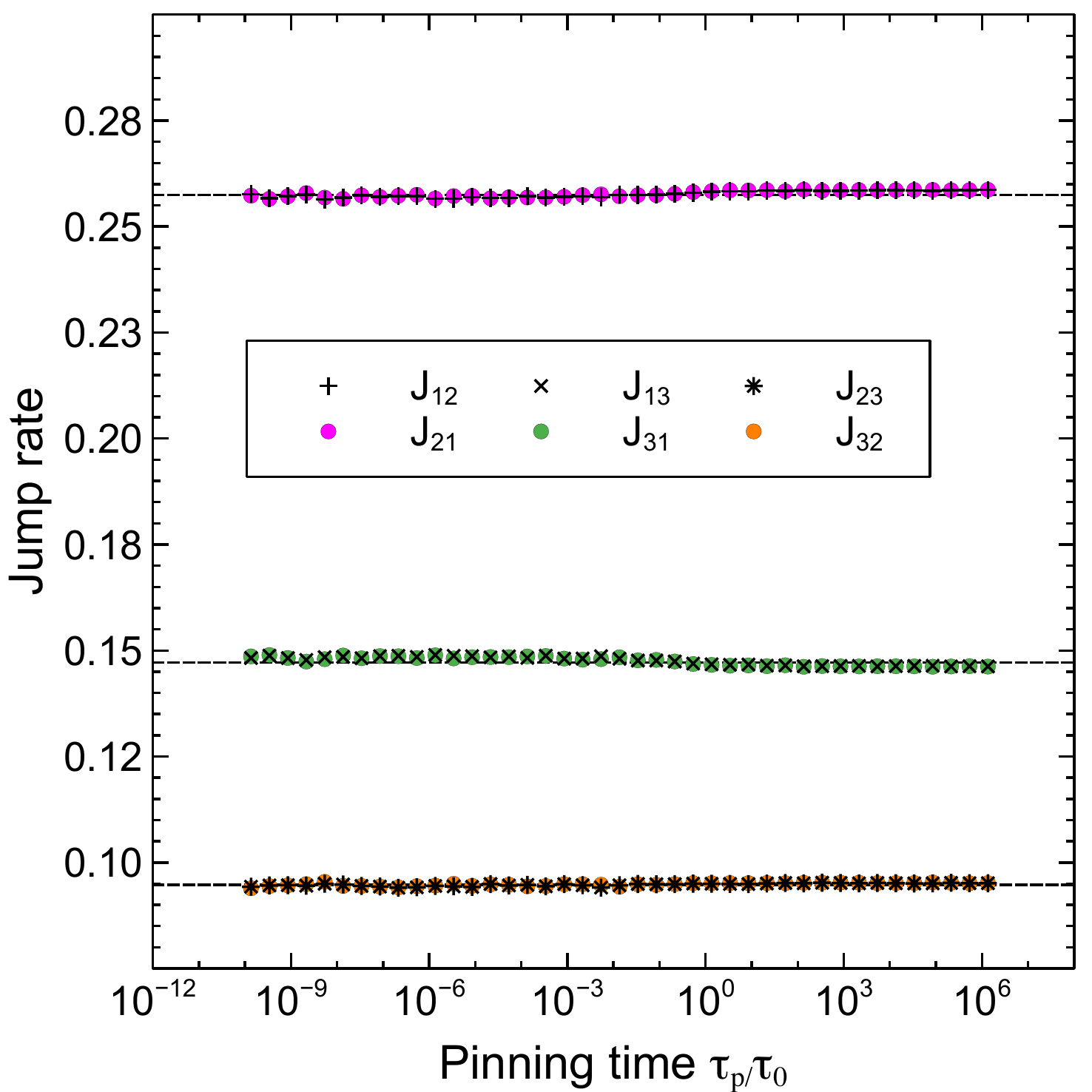}\enskip{}\enskip{}\enskip{}\textbf{(b)}\includegraphics[width=0.43\textwidth]{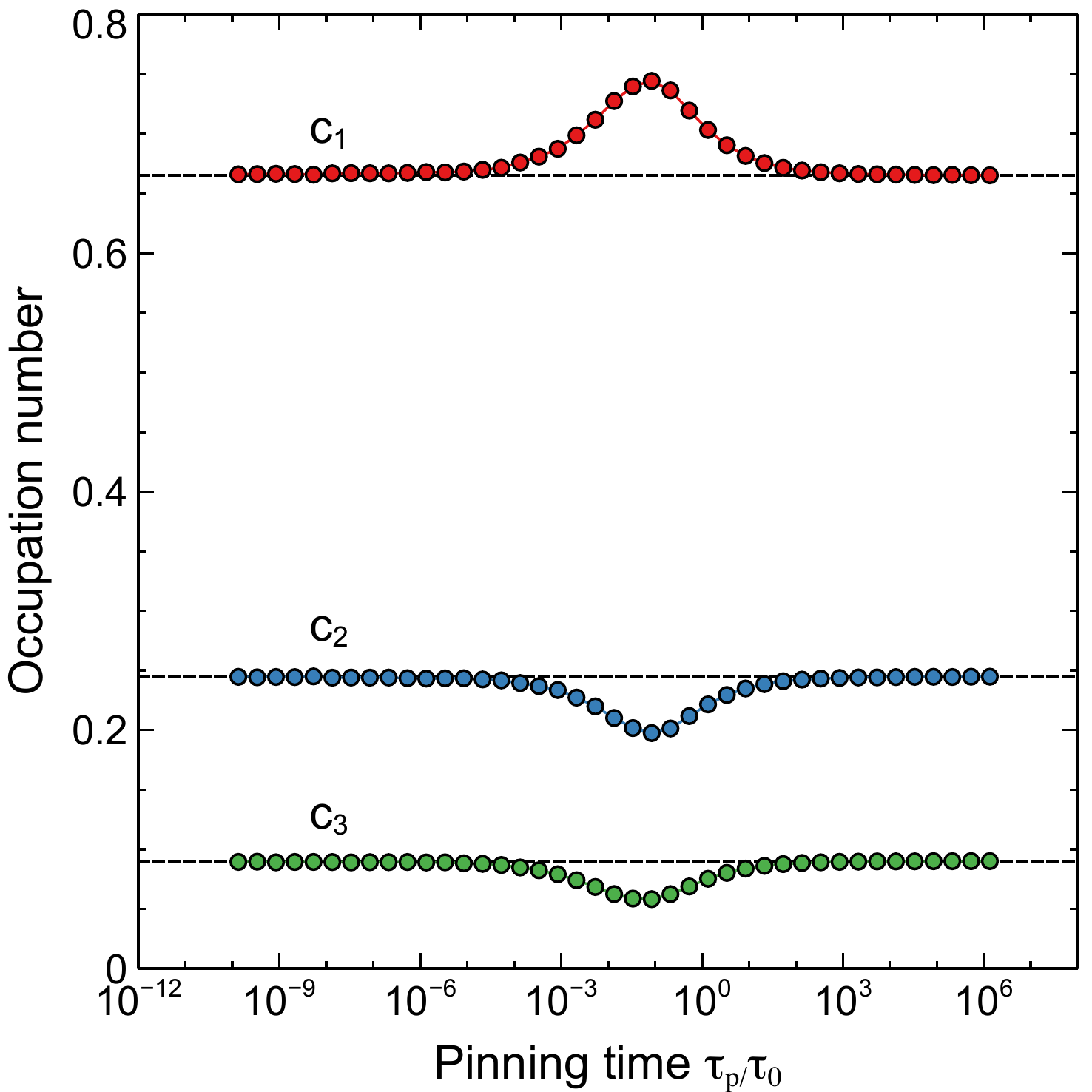}

\bigskip{}
\bigskip{}
\bigskip{}

\textbf{(c)}\includegraphics[width=0.43\textwidth]{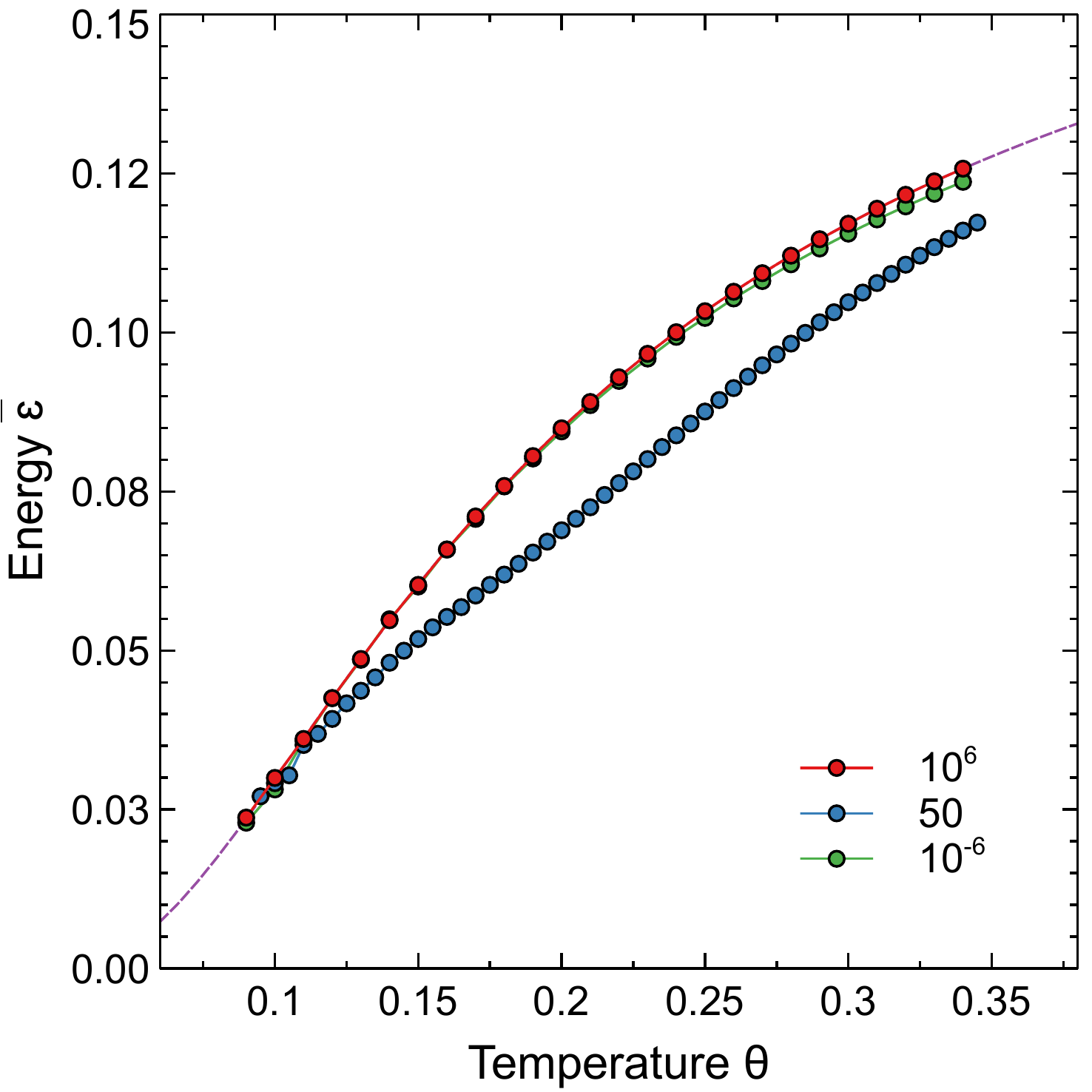}\enskip{}\enskip{}\enskip{}\textbf{(d)}\includegraphics[width=0.43\textwidth]{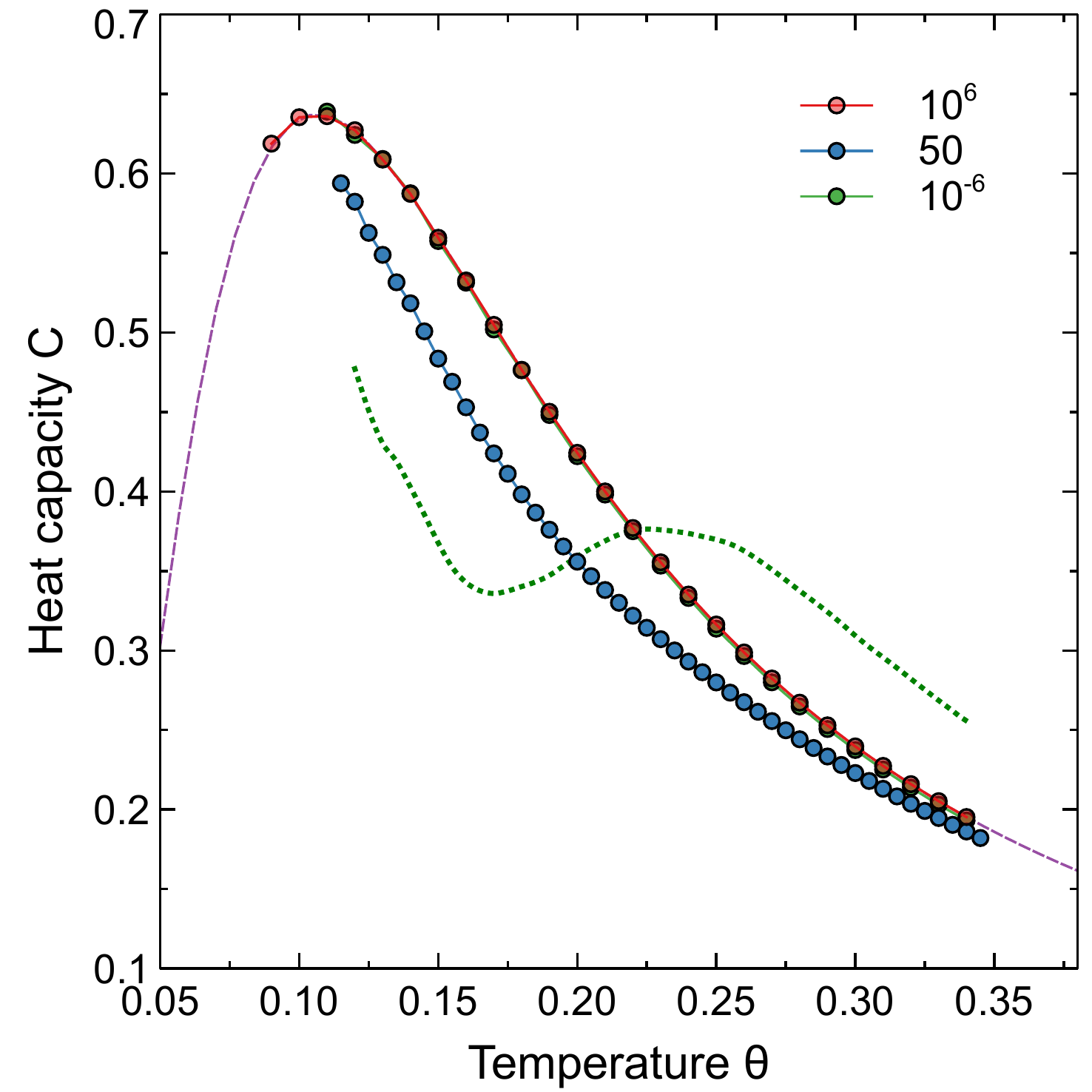}

\caption{Results of KMC simulations for a three-level system with the pinning
strength of $\alpha=1.5$. (a) Jump rates at the temperature of $\theta=0.2$
as a function of reduced pinning time $\tau_{p}/\tau_{0}$ ($\tau_{0}=74$
being the residence time at this temperature). (b) State occupation
probabilities as a function of $\tau_{p}/\tau_{0}$ at $\theta=0.2$.
(c) Expectation value of the system energy as a function of temperature
for three values of the pinning time $\tau_{p}$ shown in the legend.
(d) Heat capacity as a function of temperature for the same three
pinning times. The points were computed from the fluctuation formula
(\ref{eq:s15}). The dashed line represents the true heat capacity
$C=d\bar{\varepsilon}/d\theta$ obtained by numerical differentiation.
\label{fig:A1}}
\end{figure}

For any choice of $\theta$ and $\tau_{p}$, we find that the simulations
converge to the same steady state regardless of the initial state.
The detailed balance condition (\ref{eq:s5}) is satisfied within
the statistical scatter of the results. This is illustrated in Fig.~\ref{fig:A1}(a)
for a system with energy levels $u_{1}=0$, $u_{2}=0.2$, and $u_{3}=0.4$
at the temperature of $\theta=0.2$. The plot shows that the detailed
balance is followed in the unpinned, fully pinned, as well as the
active pinning regimes with the same set of jump rates $J_{ij}$ independent
of $\tau_{p}$. In other words, the pinning does not affect the steady-state
jump rates between the states.

As expected, the steady-state occupation probabilities in the unpinned
($\tau_{p}\gg\tau_{0}$) and fully pinned ($\tau_{p}\ll\tau_{0}$)
regimes follow the Boltzmann distribution (Fig.~\ref{fig:A1}(b)).
However, in the active pinning regime ($\tau_{p}\approx\tau_{0}$)
they significantly deviate from the $c_{i}^{B}$ values. Such deviations
are unsurprising and could be anticipated from the following considerations.
When the system is in a low-energy state, the jump barriers to other
states are high and the system spends a long time trying to escape.
The pinning process then has enough time to raise the barriers further,
making the residence time longer and thus $c_{i}$ larger than in
the absence of pinning. When the system is in a high-energy state,
the surrounding barriers are low and the system has a good chance
to escape before any significant pinning can occur. Thus, one can
expect that the pinning should shift the occupation probabilities
toward lower-energy states compared with $c_{i}^{B}$. This trend
is indeed observed in Fig.~\ref{fig:A1}(b), where $c_{1}$ exhibits
a local maximum while $c_{2}$ and $c_{3}$ local minima when $\tau_{p}$
becomes comparable to $\tau_{0}$. 

The non-Boltzmann shift of the occupation probabilities $c_{i}$ towards
lower-energy states causes a negative deviation of the system energy
$\bar{\varepsilon}=\sum_{i}c_{i}u_{i}$ from the Boltzmann energy
$\varepsilon_{B}$. Figure \ref{fig:A1}(c) shows the temperature
dependence $\varepsilon(\theta)$ along with Boltzmann's energy $\varepsilon_{B}(\theta)$
computed from Eq.(\ref{eq:s8}). The pinning times of $\tau_{p}=10^{6}$
and $\tau_{p}=10^{-6}$ represent the unpinned and fully pinned situations,
respectively. In both cases, the system energy is the same and close
to $\varepsilon_{B}(\theta)$. Accordingly, the heat capacity computed
from the canonical fluctuation relation 
\begin{equation}
C(\theta)=\dfrac{1}{\theta^{2}}\left(\overline{\varepsilon^{2}}-\bar{\varepsilon}^{2}\right)\label{eq:s15}
\end{equation}
is also the same in both cases and close to $C_{B}(\theta)$ given
by Eq.(\ref{eq:s9}) (Fig.~\ref{fig:A1}(d)). The active pinning
effect is represented by $\tau_{p}=50$. In this case, the pinning
evolves with temperature from weak at $\theta=0.1$ ($\tau_{p}/\tau_{0}=220$)
to strong at $\theta=0.35$ ($\tau_{p}/\tau_{0}=0.174$). The most
active pinning occurs at $\theta=0.215$ ($\tau_{p}/\tau_{0}=1$).
As expected, in the weak and strong pinning cases, the system energy
tends to $\varepsilon_{B}(\theta)$ while the heat capacity approaches
$C_{B}(\theta)$. In between, the energy exhibits the expected downward
deviation from $\varepsilon_{B}(\theta)$. Accordingly, the true heat
capacity computed directly from its definition, $C=d\bar{\varepsilon}/d\theta$,
deviates from $C_{B}(\theta)$. It also deviates from the heat capacity
computed from the fluctuation formula (\ref{eq:s15}). These deviations
show that the system no longer follows the canonical fluctuation theory
\citep{Landau-Lifshitz-Stat-phys,Mishin:2015ab} underlying Eq.(\ref{eq:s15}). 

To summarize, this simple model demonstrates several features of a
system subject to the pinning effect. In KMC simulations based on
this model, the system reaches a unique steady state independent of
the initial condition. The steady-state jump rates $J_{ij}$ are unaffected
by the pinning and follow the detailed balance condition. However,
the occupation probabilities of the states no longer follow the Boltzmann
distribution. The equilibrium fluctuations do not follow the canonical
relations. In particular, the energy fluctuation formula (\ref{eq:s15})
does not yield the correct heat capacity of the system. 

In the three-level model considered here, the pinning causes negative
deviations of the system energy from Eq.(\ref{eq:s8}) based on the
Boltzmann distribution. We cannot exclude, however, that the sign
of this deviation can be different in more complex systems with highly
degenerate energy levels. 

%\bibliographystyle{/Users/ymishin/YURI/Bibliography/JdePhys}
%\bibliography{/Users/ymishin/YURI/Bibliography/literat}

\begin{thebibliography}{10}

\bibitem{Balluffi95}
{\sc A.~P. Sutton and R.~W. Balluffi}.
\newblock {\em Interfaces in Crystalline Materials\/}.
\newblock Clarendon Press, Oxford,  (1995).

\bibitem{Cahn-1962}
{\sc J.~W. Cahn}.
\newblock The impurity-drag effect in grain boundary motion.
\newblock {\em Acta Metall.\/} {\bf 10} (1962) 789--798.

\bibitem{Lucke-Stuwe-1963}
{\sc K.~L{\"u}cke and H.~P. St{\"u}we}.
\newblock On the theory of grain boundary motion.
\newblock In: L.~Himmel, ed., {\em Recovery and Recrystallization of Metals\/}.
  Interscience Publishers, New York, 1963  171--210.

\bibitem{Lucke:1971aa}
{\sc K.~L{\"u}cke and H.~P. St{\"u}we}.
\newblock On the theory of impurity controlled grain boundary motion.
\newblock {\em Acta Metall.\/} {\bf 19} (1971) 1087--1099.

\bibitem{Wang03}
{\sc N.~Ma, S.~A. Dregia and Y.~Wang}.
\newblock Solute segregation transition and drag force on grain boundaries.
\newblock {\em Acta Mater.\/} {\bf 51} (2003) 3687--3700.

\bibitem{Li:2009aa}
{\sc J.~Li, J.~Wang and G.~Yang}.
\newblock Phase field modeling of grain boundary migration with solute drag.
\newblock {\em Acta Mater.\/} {\bf 57} (2009) 2108--2120.

\bibitem{Shahandeh:2012aa}
{\sc S.~Shahandeh, M.~Greenwood and M.~Militzer}.
\newblock Friction pressure method for simulating solute drag and particle
  pinning in a multiphase-field model.
\newblock {\em Model. Simul. Mater. Sci. Eng.\/} {\bf 20} (2012) 065008.

\bibitem{Gronhagen:2007aa}
{\sc K.~Gr{\"o}nhagen and J.~Agren}.
\newblock Grain-boundary segregation and dynamic solute drag theory --- {A}
  phase-field approach.
\newblock {\em Acta Mater.\/} {\bf 55} (2007) 955--960.

\bibitem{Abdeljawad:2017aa}
{\sc F.~Abdeljawad, P.~Lu, N.~Argibay, B.~G. Clark, B.~L. Boyce and S.~M.
  Foiles}.
\newblock Grain boundary segregation in immiscible nanocrystalline alloys.
\newblock {\em Acta Mater.\/} {\bf 126} (2017) 528--539.

\bibitem{Alkayyali:2021uo}
{\sc M.~Alkayyali and F.~Abdeljawad}.
\newblock Grain boundary solute drag model in regular solution alloys.
\newblock {\em Physical Review Letters\/} {\bf 127} (2021) 175503.

\bibitem{Greenwood:2012aa}
{\sc M.~Greenwood, C.~Sinclair and M.~Militzer}.
\newblock Phase field crystal model of solute drag.
\newblock {\em Acta Mater.\/} {\bf 60} (2012) 5752--5761.

\bibitem{Mendelev02a}
{\sc M.~I. Mendelev and D.~J. Srolovitz}.
\newblock Impurity effects on grain boundary migration.
\newblock {\em Model. Simul. Mater. Sci. Eng.\/} {\bf 10} (2002) R79--R109.

\bibitem{Sun:2014aa}
{\sc H.~Sun and C.~Deng}.
\newblock Direct quantification of solute effects on grain boundary motion by
  atomistic simulations.
\newblock {\em Comp. Mater. Sci.\/} {\bf 93} (2014) 137--143.

\bibitem{Wicaksono:2013aa}
{\sc A.~T. Wicaksono, C.~W. Sinclair and M.~Militzer}.
\newblock A three-dimensional atomistic kinetic {Monte Carlo} study of dynamic
  solute-interface interaction.
\newblock {\em Model. Simul. Mater. Sci. Eng.\/} {\bf 21} (2013) 085010.

\bibitem{Mendelev:2001aa}
{\sc M.~I. Mendelev, D.~J. Srolovitz and W.~E}.
\newblock Grain-boundary migration in the presence of diffusing impurities:
  simulations and analytical models.
\newblock {\em Philos. Mag.\/} {\bf 81} (2001) 2243--2269.

\bibitem{Rahman:2016aa}
{\sc M.~J. Rahman, H.~S. Zurob and J.~J. Hoyt}.
\newblock Molecular dynamics study of solute pinning effects on grain boundary
  migration in the aluminum magnesium alloy system.
\newblock {\em Metall. Mater. Trans. A\/} {\bf 47} (2016) 1889--1897.

\bibitem{Kim:2008aa}
{\sc S.~G. Kim and Y.~B. Park}.
\newblock Grain boundary segregation, solute drag and abnormal grain growth.
\newblock {\em Acta Mater.\/} {\bf 56} (2008) 3739--3753.

\bibitem{Koju:2020aa}
{\sc R.~Koju and Y.~Mishin}.
\newblock Direct atomistic modeling of solute drag by moving grain boundaries.
\newblock {\em Acta Mater.\/} {\bf 198} (2020) 111--120.

\bibitem{Koju:2021aa}
{\sc R.~K. Koju and Y.~Mishin}.
\newblock The role of grain boundary diffusion in the solute drag effect.
\newblock {\em Nanomaterials\/} {\bf 11} (2021) 2348.

\bibitem{Mishin_2023_RW_part_II}
{\sc Y.~Mishin}.
\newblock submitte as Part II of this work.

\bibitem{Vineyard:1957vo}
{\sc G.~H. Vineyard}.
\newblock Frequency factors and isotope effects in solid state rate processes.
\newblock {\em Journal of Physics and Chemistry of Solids\/} {\bf 3} (1957)
  121--127.

\bibitem{Cahn:2001wh}
{\sc J.~W. Cahn and F.~R.~N. Nabarro}.
\newblock Thermal activation under shear.
\newblock {\em Philosophical Magazine A\/} {\bf 81} (2001) 1409--1426.

\bibitem{Cottrell:2002ub}
{\sc A.~H. Cottrell}.
\newblock Thermally activated plastic glide.
\newblock {\em Philosophical Magazine Letters\/} {\bf 82} (2002) 65--70.

\bibitem{Ivanov08a}
{\sc V.~A. Ivanov and Y.~Mishin}.
\newblock Dynamics of grain boundary motion coupled to shear deformation: An
  analytical model and its verification by molecular dynamics.
\newblock {\em Phys. Rev. {\rm B}\/} {\bf 78} (2008) 064106.

\bibitem{Chachamovitz:2018vm}
{\sc D.~Chachamovitz and D.~Mordehai}.
\newblock The stress-dependent activation parameters for dislocation nucleation
  in molybdenum nanoparticles.
\newblock {\em Scientific Reports\/} {\bf 8} (2018) 3915.

\bibitem{Hirth}
{\sc J.~P. Hirth and J.~Lothe}.
\newblock {\em Theory of Dislocations\/}.
\newblock Wiley, New York, second edition,  (1982).

\bibitem{Cottrell:1948aa}
{\sc A.~H. Cottrell}.
\newblock Effect of solute atoms on behavior of dislocations.
\newblock In: {\em Report of a Conference on Strength of Solids\/}, Lodon, UK,
  (1948). The Physical Society, 1948  30--38.

\bibitem{Cottrell_Bilby}
{\sc A.~Cottrell and B.~A. Bilby}.
\newblock Dislocation theory of yielding and strain aging of iron.
\newblock {\em Proc. Phys. Soc. London\/} {\bf 62} (1949) 49--62.

\bibitem{Cottrell:1953aa}
{\sc A.~H. Cottrell}.
\newblock {\em Dislocations and plastic flow in crystals\/}.
\newblock Clarendon Press, Oxford,  (1953).

\bibitem{Chari:1994aa}
{\sc M.~K. Chari}.
\newblock On reversible semi-Markov processes.
\newblock {\em Operations Research Letters\/} {\bf 15} (1994) 157--161.

\bibitem{Yu:2010aa}
{\sc S.-Z. Yu}.
\newblock Hidden semi-Markov models.
\newblock {\em Artificial Intelligence\/} {\bf 174} (2010) 215--243.

\bibitem{Wang:2007aa}
{\sc H.~Wang and H.~Qian}.
\newblock On detailed balance and reversibility of semi-{Markov} processes and
  single-molecule enzyme kinetics.
\newblock {\em Journal of Mathematical Physics\/} {\bf 48} (2007) 013303.

\bibitem{Maes:2009aa}
{\sc C.~Maes, K.~Neto{\v c}n{\'y} and B.~Wynants}.
\newblock Dynamical fluctuations for semi-{Markov} processes.
\newblock {\em Journal of Physics A: Mathematical and Theoretical\/} {\bf 42}
  (2009) 365002.

\bibitem{Landau-Lifshitz-Stat-phys}
{\sc L.~D. Landau and E.~M. Lifshitz}.
\newblock {\em Statistical Physics, Part I\/}, volume~5 of {\em Course of
  Theoretical Physics\/}.
\newblock Butterworth-Heinemann, Oxford, third edition,  (2000).

\bibitem{Mishin:2015ab}
{\sc Y.~Mishin}.
\newblock Thermodynamic theory of equilibrium fluctuations.
\newblock {\em Annals of Physics\/} {\bf 363} (2015) 48--97.

\bibitem{Manousiouthakis:1999wn}
{\sc V.~I. Manousiouthakis and M.~W. Deem}.
\newblock Strict detailed balance is unnecessary in {Monte Carlo} simulation.
\newblock {\em The Journal of Chemical Physics\/} {\bf 110} (1999) 2753--2756.

\end{thebibliography}

\newpage\clearpage{}


\clearpage
\end{document}